\documentclass[aps,prl,superscriptaddress,twocolumn,showpacs]{revtex4-1} 

\usepackage{amsmath,amssymb,mathrsfs}
\usepackage{latexsym}
\usepackage{graphicx} 
\usepackage{epstopdf}
\usepackage{graphicx,epstopdf,color}
\usepackage{amsfonts}
\usepackage{bm}

\newcommand{\1}{\hspace*{-1pt}}

\newcommand{\3}{\hspace*{-3pt}}

\newcommand{\bs}[1]{\boldsymbol{#1}}

\begin{document}

\title{Anisotropic susceptibilities in the honeycomb Kitaev system $\bs{\alpha}$-RuCl$_3$}

\author{P. Lampen-Kelley}
\affiliation{Department of Materials Science and Engineering, University of Tennessee, Knoxville, TN 37996, U.S.A.}
\affiliation{Materials Science and Technology Division, Oak Ridge National Laboratory, Oak Ridge, TN, 37831, U.S.A.}
\author{S. Rachel}
\affiliation{School of Physics, University of Melbourne, Parkville, VIC 3010, Australia}
\affiliation{Institut f\"ur Theoretische Physik, Technische Universit\"at Dresden, 01062 Dresden, Germany}
\author{J. Reuther}
\affiliation{Dahlem Center for Complex Quantum Systems and Fachbereich Physik, Freie Universit\"at Berlin, 14195 Berlin, Germany}
\affiliation{Helmholtz-Zentrum Berlin f\"ur Materialien und Energie, D-14109 Berlin, Germany}
\author{J.-Q. Yan}
\affiliation{Materials Science and Technology Division, Oak Ridge National Laboratory, Oak Ridge, TN, 37831, U.S.A.}
\author{A. Banerjee}
\affiliation{Neutron Scattering Division, Oak Ridge National Laboratory, Oak Ridge, TN 37831, U.S.A.}
\author{C.A. Bridges}
\affiliation{Chemical Sciences Division, Oak Ridge National Laboratory, Oak Ridge, TN 37831, U.S.A.}
\author{H.B. Cao}
\affiliation{Neutron Scattering Division, Oak Ridge National Laboratory, Oak Ridge, TN 37831, U.S.A.}
\author{S.E. Nagler}
\affiliation{Neutron Scattering Division, Oak Ridge National Laboratory, Oak Ridge, TN 37831, U.S.A.}
\author{D. Mandrus}
\affiliation{Department of Materials Science and Engineering, University of Tennessee, Knoxville, TN 37996, U.S.A.}
\affiliation{Materials Science and Technology Division, Oak Ridge National Laboratory, Oak Ridge, TN, 37831, U.S.A.}
 \pagestyle{plain}

\begin{abstract}
The magnetic insulator $\alpha$-RuCl$_3$ is a promising candidate to realize Kitaev interactions on a quasi-2D honeycomb lattice. We perform extensive susceptibility measurements on single crystals of $\alpha$-RuCl$_3$, including angle-dependence of the in-plane longitudinal and transverse susceptibilities, which reveal a unidirectional anisotropy within the honeycomb plane. By comparing the experimental results to a high-temperature expansion of a Kitaev-Heisenberg-$\Gamma$ spin Hamiltonian with bond-anisotropy, we find excellent agreement with the observed phase shift and periodicity of the angle-resolved susceptibilities. 
Within this model, we show that the pronounced difference between in-plane and out-of-plane susceptibilities as well as the finite transverse susceptibility are rooted in strong symmetric off-diagonal $\Gamma$ spin exchange. The $\Gamma$ couplings and relationships between other terms in the model Hamiltonian are quantified by extracting relevant Curie-Weiss intercepts from the experimental data.
\end{abstract}

\date{\today}

\maketitle

\paragraph{Introduction.}

Quantum spin liquids are exotic states of matter in which the formation of conventional long-range order is avoided down to the lowest temperatures due to strong quantum fluctuations \cite{savary_quantum_2017, zhou_quantum_2017}. A number of frustrated magnets are promising candidates to host quantum spin liquid ground states \cite{balents_spin_2010}, however both the theoretical prediction and the experimental observation of such spin liquids are notoriously difficult, since clear identifying signatures are uncommon in the absence of any order.  A notable exception is the Kitaev honeycomb model, a spin Hamiltonian with an exactly solvable spin liquid ground state \cite{kitaev_anyons_2006}. The exact solvability of the model allows for the extraction of insights and details which can be very difficult to determine for more generic systems \cite{hermanns_physics_2018}.

Consequently, there has been considerable effort over the past several years to identify materials which realize Kitaev spin exchange \cite{hermanns_physics_2018,jackeli_mott_2009,trebst_kitaev_2017,kimchi_kitaev-heisenberg_2014,aczel_highly_2016,yamada_designing_2017,liu_pseudospin_2017}. Potential manifestations of the 2D Kitaev model are found in the layered honeycomb magnetic insulators A$_2$IrO$_3$ (A=Na, Li, Cu) \cite{jackeli_mott_2009,singh_relevance_2012,hwan_chun_direct_2015, abramchuk_cu2iro3:_2017} and $\alpha$-RuCl$_3$ \cite{plumb_$ensuremathalpha-mathrmrucl_3$:_2014,koitzsch_$j_mathrmeff$_2016,sandilands_scattering_2015,banerjee_neutron_2017}. Kitaev interactions in these systems are accompanied by more conventional spin exchange, leading to long-range magnetic order at low temperatures \cite{ye_direct_2012,williams_incommensurate_2016,sears_magnetic_2015,cao_low-temperature_2016,winter_challenges_2016,rau_generic_2014,rousochatzakis_phase_2015} with the exception of Cu$_2$IrO$_3$ which exhibits a short-range magnetic order \cite{abramchuk_cu2iro3:_2017}. 
Despite the rapidly increasing interest in these materials, the effective spin Hamiltonian that best captures the experimental results remains controversial - see discussion in \cite{janssen_magnetization_2017} and references therein.

\begin{figure}[t]
\centering
\includegraphics[width=0.8\linewidth]{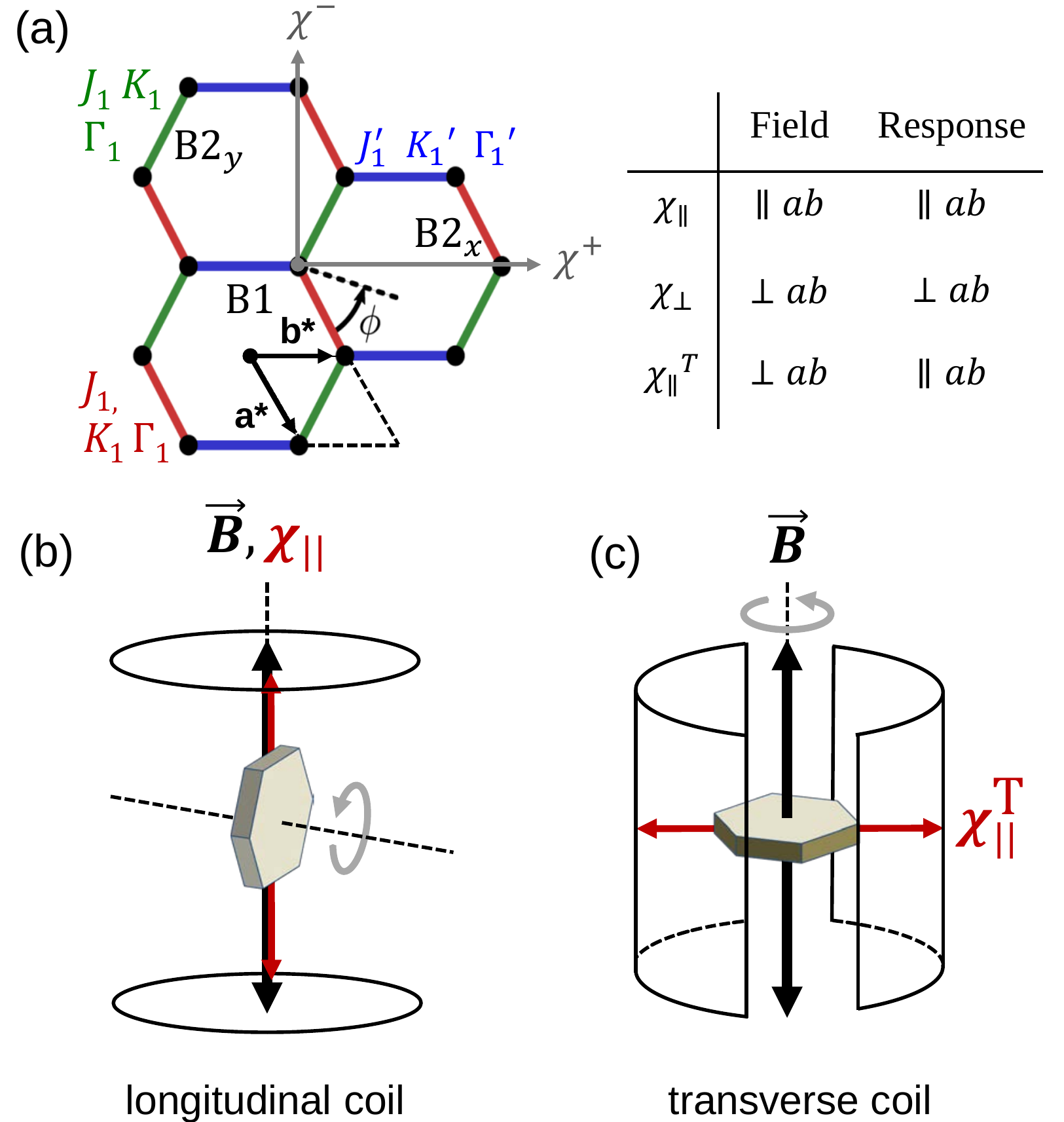}

\caption{(a) Definition of the in-plane angle $\phi$, the three Ru-Ru bonds B1, B2$_x$, and B2$_y$, and notation for various measurement configurations. In a trigonal setting, the bonds are parallel to the (0,1,0), (1,0,0), and ($\bar{1}$,1,0) reciprocal lattice vectors, respectively. Sketch of the geometry for in-plane angle-resolved susceptibility measurements in (b) longitudinal SQUID coils and (c) transverse SQUID coils. The angle dependence is mapped out as the sample is rotated 360$^\circ$ about a fixed axis perpendicular or parallel to the direction of the applied field $\vec{B}$, respectively, yielding $\chi_{||}(\phi)$ and $\chi_\parallel^\text{T}(\phi)$.   }
\label{fig:lattice}
\end{figure}

A marked anisotropy between the magnetic susceptibilities measured with a magnetic field applied parallel $\chi_{||}$ or perpendicular $\chi_{\perp}$ to the honeycomb plane has been reported in $\alpha$-RuCl$_3$ \cite{sears_magnetic_2015,majumder_anisotropic_2015,kubota_successive_2015} and A$_2$IrO$_3$ \cite{singh_antiferromagnetic_2010,freund_single_2016}. However, a systematic explanation for this phenomenon in terms of microscopic exchange couplings has not yet been given. Moreover, experimental results which involve a magnetic field applied parallel to the honeycomb plane depend on the in-plane angle of the applied field \cite{leahy_anomalous_2017,little_antiferromagnetic_2017,ponomaryov_unconventional_2017}. Motivated by these observations, we perform extensive susceptibility measurements on single crystals of $\alpha$-RuCl$_3$. The longitudinal and transverse susceptibilities as a function of angle within the honeycomb plane are compared to a high-temperature expansion of the magnetic susceptibility tensor for a bond-anisotropic Kitaev-Heisenberg-$\Gamma$ model. Given the excellent agreement between the model and experimental results, we suggest mechanisms for the observed anisotropies and extract quantitative relationships between terms in the model Hamiltonian.


\begin{figure}[]
\centering
\includegraphics[width=1.03\linewidth]{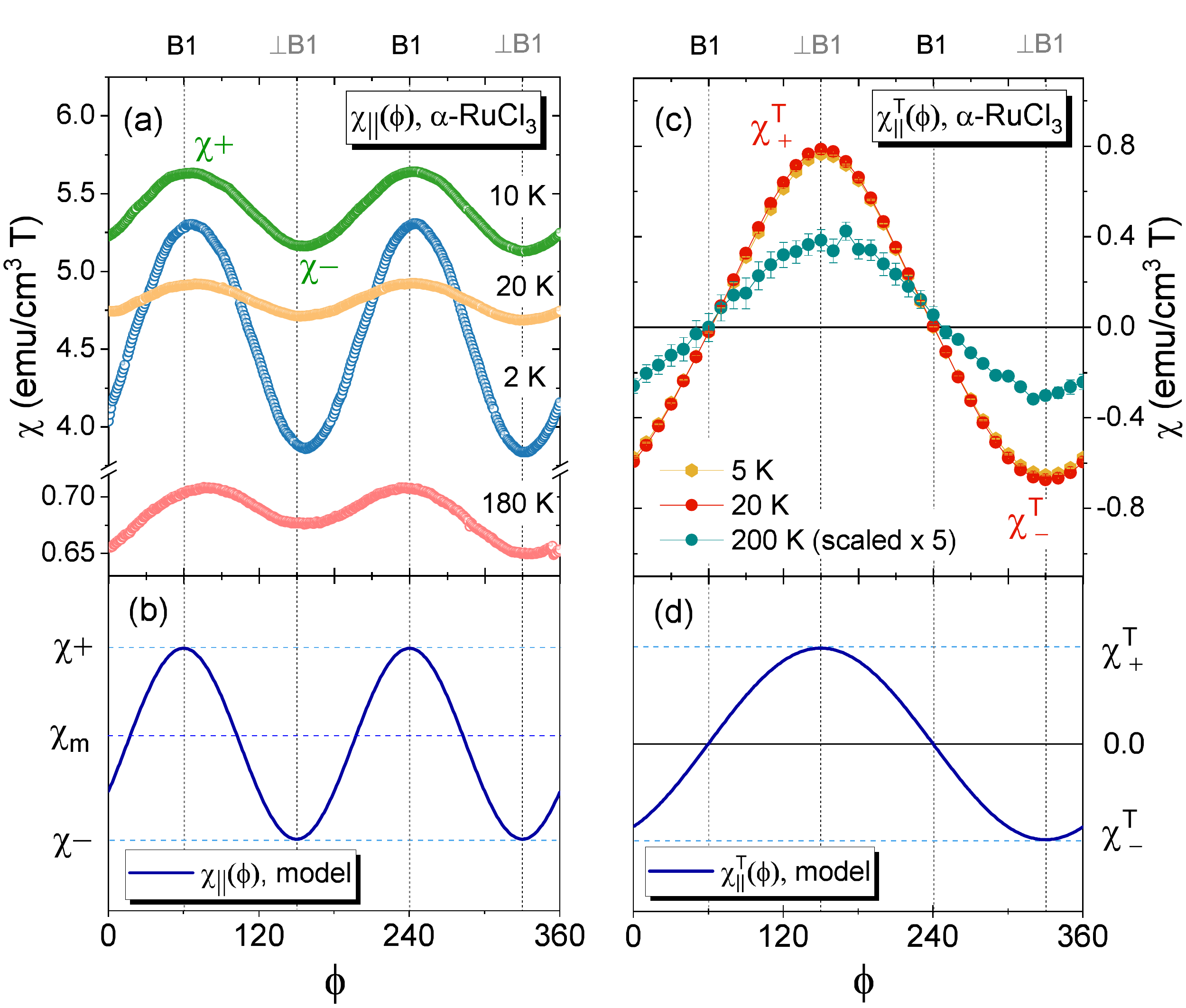}
\caption{(a) Angle-resolved longitudinal susceptibility $\chi_\parallel(\phi)$ of a single crystal of $\alpha$-RuCl$_3$ as the direction of magnetic field ($B = 0.1$ T) is varied within the honeycomb plane. $\phi$ is the in-plane angle between \textbf{a*} and the measurement direction. A diamagnetic contribution from the rotation stage is subtracted from the presented data. (b) Theoretical oscillation of $\chi_\parallel(\phi)$ predicted for a bond-anisotropic Kitaev-Heisenberg-$\Gamma$ model, see Eq.~(\ref{chi_parallel}). (c) Angle-resolved transverse susceptibility $\chi_\parallel^\text{T}(\phi)$ of a single crystal of $\alpha$-RuCl$_3$ as a function of the in-plane angle $\phi$ with $B = 1$ T applied perpendicular to the ab plane. The 200 K data are scaled by a factor of 5 to facilitate viewing on the same axes. (d) Theoretical oscillation of $\chi_\parallel^\text{T}(\phi)$ resulting from Eq.~(\ref{off}). 
The location of the anisotropic bond (B1) is marked on the upper horizontal axis.}
\label{fig:inplane_osc}
\end{figure}

\paragraph{Experimental details.} Single crystals of $\alpha$-RuCl$_3$ were prepared using a vapor transport technique ~\cite{banerjee_neutron_2017}, and crystallographic directions were identified prior to susceptibility measurements via Laue diffraction. Angle-resolved in-plane longitudinal  $\chi_{||} (\phi)$ and transverse  $\chi_\parallel^\text{T}(\phi)$ magnetic susceptibilities were measured using commercial SQUID magnetometers (Quantum Design) \cite{QDnote}, where $\phi$ is the angle between the measurement direction and \textbf{a*} (for simplicity, we adopt a trigonal notation - see Fig. 1a). Figure 1(b) depicts the standard longitudinal measurement geometry, in which the magnetic field $\vec{B}$ and measurement axes coincide and the diagonal elements $\chi^{\mu \mu}$ of the susceptibility tensor are determined. Off-diagonal elements $\chi^{\mu \mu'}$ are accessed in a transverse SQUID geometry where susceptibility is measured along an axis perpendicular to $\vec{B}$  (Fig. 1c). To eliminate uncompensated longitudinal moment in the transverse pick-up coils, the raw SQUID voltage was decomposed into even and odd signals before fitting the even component to an appropriate response function to extract the transverse moment at each condition \cite{QDnote, thompson_field_2010}.

\paragraph{Oscillating susceptibility.} Figure \,\ref{fig:inplane_osc}(a) shows the longitudinal susceptibility of an $\alpha$-RuCl$_3$ single crystal as the direction of the magnetic field varies within the $ab$ plane (see Fig. 1b). Clear oscillations in the magnitude of the in-plane susceptibility $\chi_\parallel(\phi)$ with $\pi$-periodicity are observed both below and above the zigzag magnetic ordering transition at $T_\text{N} \simeq 7$ K, suggesting that the appearance of in-plane magnetic anisotropy is not tied to long-range order. The maxima (minima) of $\chi_\parallel$ occur at $\phi = 60^\circ$ and $240^\circ$ ($\phi=150^\circ$ and $330^\circ$), corresponding to magnetic field parallel (perpendicular) to one of the Ru-Ru bond directions. This inequivalent bond is referred to hereafter as B1 (Fig. 1a). Oscillations in $\chi_\parallel$ with $\pi$-periodicity persist for $T \gg T_\text{N}$ even as the mean value $\chi_m$ decays with the overall susceptibility at high temperatures. 

Oscillations are also observed in the in-plane transverse susceptibility $\chi_\parallel^\text{T}$, where the magnetic field is applied along (0,0,1). Figure 2(c) shows the $\phi$-dependence of $\chi_\parallel^\text{T}$ as the crystal was rotated about a vertical axis coinciding with the field direction. Both below and above $T_\text{N}$, the susceptibility shows a well-defined oscillation about zero with a $2\pi$ period. The absolute maxima (nodes) of the oscillation occur perpendicular (parallel) to  the inequivalent B1 bond at $\phi=150^\circ$ and $330^\circ$ ($\phi = 60^\circ$ and $240^\circ$).

A number of space groups, distinguished primarily by the stacking sequence of van der Waals-coupled honeycomb layers, have been proposed for $\alpha$-RuCl$_3$ \cite{kim_crystal_2016}. Most recently, a structural transition from high-temperature monoclinic $C2/m$ to trigonal $R\overline{3}$ was reported at $T\simeq 150$ K \cite{park_emergence_2016}. Our analysis below relies on a high-temperature model expansion, and thus a quantitative comparison to the model is made within the monoclinic phase. The sinlgle-domain monoclinic structure of the sample for which data is presented in Figs. 2-4 was confirmed directly by single crystal neutron diffraction at $T > 150$ K using the HB-3A beamline at the High Flux Isotope Reactor, Oak Ridge National Laboratory. The small inequivalence in one of the Ru-Ru bond lengths \cite{cao_low-temperature_2016} provides a natural explanation for the observation of a unique magnetically easy direction. We note that the $\pi$-period oscillation observed at all temperatures in this work, as well as in-plane anisotropy reported in a recent THz study \cite{little_antiferromagnetic_2017}, appear to be incompatible with a low-temperature trigonal point group. These results suggest a deviation from an ideal $R\overline{3}$ structure, which may be related to strain induced at the structural transition.

\paragraph{Model and high-temperature expansion.} To model the observed behavior we consider a variant of an anisotropic Kitaev-Heisenberg-$\Gamma$ Hamiltonian with nearest neighbor Heisenberg exchange ($J_1$), Kitaev interactions ($K_1$), and nearest neighbor symmetric off-diagonal spin exchange ($\Gamma_1$). Inequivalent interactions $J_1'$, $K_1'$, and $\Gamma_1'$ are assigned to the bond direction B1 giving a Hamiltonian of the form $H=H_{\text{B1}}+H_{\text{B2}_x}+H_{\text{B2}_y}$, where
\begin{equation}
H_{\text{B1}}=\sum_{\text{B1-bonds (ij)}}J_1'\mathbf{S}_i\mathbf{S}_j+K_1'S_i^zS_j^z+\Gamma_1'(S_i^xS_j^y+S_i^yS_j^x)\;,\label{ham1}
\end{equation}
\begin{equation}
H_{\text{B2}_x}=\sum_{\text{B2$_x$-bonds (ij)}}J_1\mathbf{S}_i\mathbf{S}_j+K_1S_i^xS_j^x+\Gamma_1(S_i^yS_j^z+S_i^zS_j^y) \;,\label{ham2}
\end{equation}
and $H_{\text{B2}_y}$ follows from $H_{\text{B2}_x}$ by replacing $x\leftrightarrow y$ \cite{footnote_2}. Note that the B1 bond is symmetry-inequivalent to the two B2$_\alpha$ bonds while $H_{\text{B2}_x}$ and $H_{\text{B2}_y}$ are related by a spin rotation. Additional further neighbor couplings can be straightforwardly included (which also applies to the $J_3$ coupling which has been proposed to be
sizeable \cite{winter_models}), however here we restrict the analysis to nearest neighbor couplings for simplicity of notation \cite{footnote2}.

A high-temperature expansion of the full zero-field susceptibility tensor $\chi^{\mu\mu'}$ ($\mu,\mu'=x,y,z$) of this model up to terms $\sim T^{-2}$ yields
\begin{align}
&\chi^{\mu\mu'}(T)
=\frac{\mu_\text{B}^2N}{4k_\text{B}T}
\left(\begin{array}{ccc}
g_x^2 & 0 & 0\\[2pt]
0 & g_x^2 & 0\\[2pt]
0 & 0 & g_z^2
\end{array}\right)-\frac{\mu_\text{B}^2N}{(4k_\text{B}T)^2}\times\notag\\[10pt]
&
\times\!\left(\begin{array}{ccc}
\3g_x^2(2J_1\3+\3J_1'\3+\3K_1) & g_x^2\Gamma_1' & g_xg_z\Gamma_1\\[3pt]
g_x^2\Gamma_1' & g_x^2(2J_1\3+\3J_1'\3+\3K_1) & g_xg_z\Gamma_1\\[3pt]
g_xg_z\Gamma_1 & g_xg_z\Gamma_1 & g_z^2(2J_1\3+\3J_1'\3+\3K_1')
\end{array}\right)\notag\\[10pt]
&+\mathcal{O}(T^{-3})\;,\label{chi_tensor}
\end{align}
where $N$ denotes the total number of spins. Here we allow for a $g$-factor anisotropy of the form $g_x=g_y\neq g_z$ due to symmetry considerations. Projecting Eq.\,\eqref{chi_tensor} onto an in-plane direction
 yields an expression for the longitudinal in-plane susceptibility $\chi_\parallel(\phi)$,
\begin{align}
&\quad \chi_\parallel(\phi)=\frac{1}{6}\Big[4\chi^{xx}+2\chi^{zz}-2\chi^{xy} -4\chi^{xz} \Big.\notag\\
&\Big.+\big(\chi^{xx}-\chi^{zz}-2\chi^{xy}+2\chi^{xz}\big)\big(-\cos(2\phi)+\sqrt{3}\sin(2\phi)\big)\Big].\label{chi_parallel}
\end{align}
The harmonic oscillation described by the term $-\cos(2\phi)+\sqrt{3}\sin(2\phi)$ is illustrated in Fig.\,\ref{fig:inplane_osc}(b), which reproduces the experimentally observed periodicity of $\chi_\parallel(\phi)$. Furthermore, the location of the extrema parallel and perpendicular to a Ru-Ru bond direction is in agreement with the measured susceptibility.

The susceptibility tensor $\chi^{\mu\mu'}$ can likewise be projected to yield an expression for the transverse in-plane susceptibility $\chi_\parallel^\text{T}(\phi)$,
\begin{equation}
\chi_\parallel^\text{T}(\phi)=\frac{1}{3\sqrt{2}}(-\chi^{xx}+\chi^{zz}-\chi^{xy}+\chi^{xz})(\sin(\phi)-\sqrt{3}\cos(\phi))\;,\label{off}
\end{equation}
where the term $\sin(\phi)-\sqrt{3}\cos(\phi)$ again reproduces the measured oscillations, showing maxima (minima) at $\phi=150^\circ$ ($330^\circ$) as well as zeros at $60^\circ$ and $240^\circ$, see Fig. 2(c,d).

The results of Fig.~\ref{fig:inplane_osc} demonstrate that the anisotropic nature of the susceptibilities in $\alpha$-RuCl$_3$ is captured well by the bond-inequivalent Kitaev-Heisenberg-$\Gamma$ model described in Eqs.~(\ref{ham1}) and (\ref{ham2}). Using the high-temperature expansion in Eq.~(\ref{chi_tensor}) and assuming an isotropic $g$-factor $g_x=g_y=g_z$, which is close to the recently reported value $g_x=g_y=1.1g_z$ \cite{agrestini_electronically_2017}, a simple interpretation of the observed oscillations and the in-plane/out-of-plane anisotropy arises: The amplitude $\chi_+ - \chi_-$ of the oscillation in $\chi_\parallel(\phi)$ (where $\chi_+$ and $\chi_-$ are the maxima and minima of $\chi_\parallel(\phi)$) is proportional to the differences of the couplings on the B1 and B2$_\alpha$-bonds,
\begin{equation}
\chi_+-\chi_-\sim T^{-2}[K_1'-K_1+2(\Gamma_1'-\Gamma_1)]\;.
\end{equation}

\begin{figure}[t]
\centering
\includegraphics[width=1.0\linewidth]{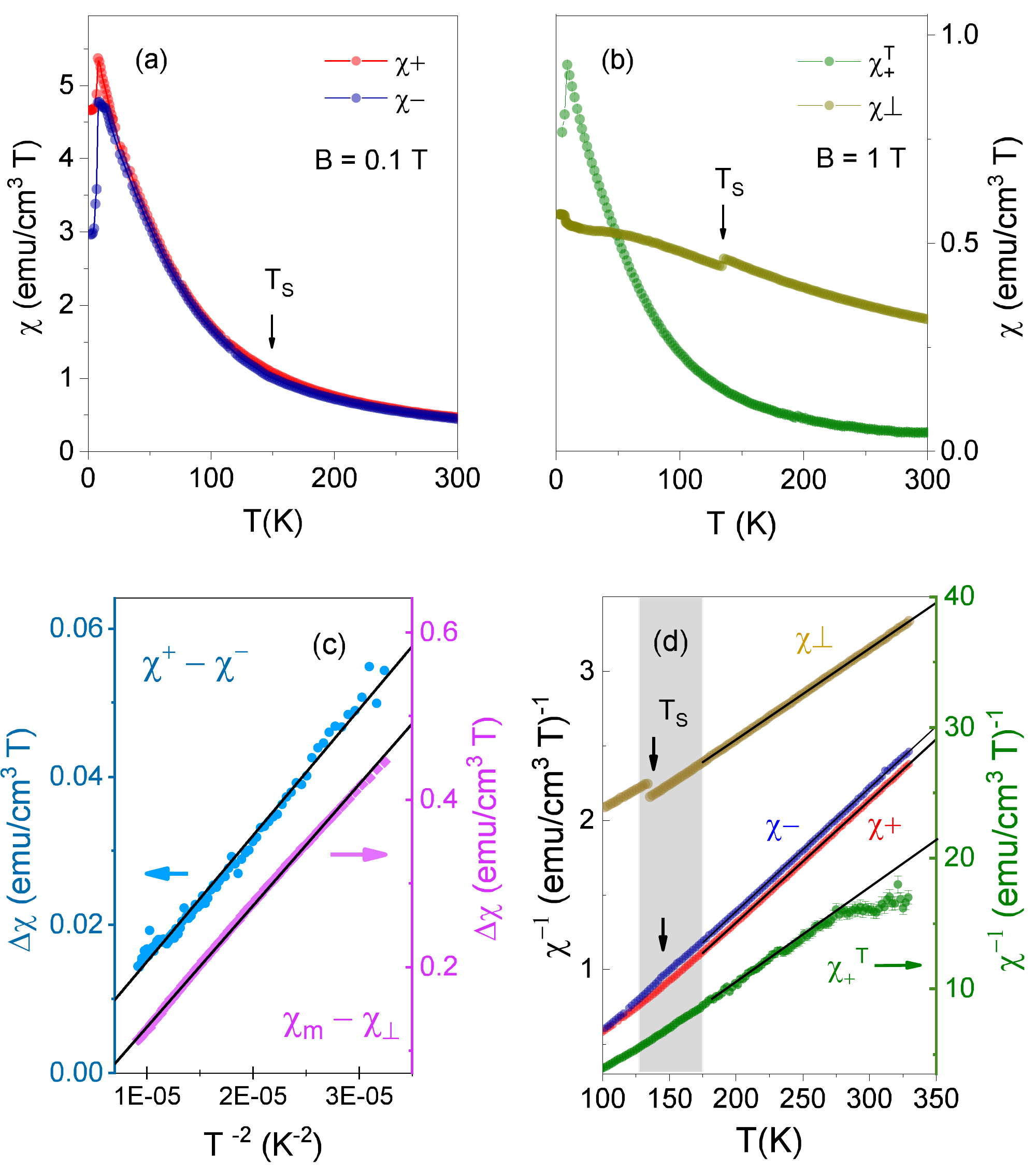}
\caption{Temperature-dependent susceptibilities at fixed angle, corrected for core diamagnetism \cite{bain_diamagnetic_2008}. (a) In-plane maximum and minimum longitudinal susceptibilities $\chi_+$ and $\chi_-$. (b) Out-of-plane longitudinal susceptibility $\chi_\perp$ and maximum transverse in-plane susceptibility $\chi_+^\text{T}$. $\chi_+$, $\chi_-$, and $\chi_+^\text{T}$ are measured in the same crystal for which angle-resolved data are shown. $\chi_\perp$ is measured on a second sample with greater thickness in the $c$ direction. (c) Differences $\chi_+ - \chi_-$ and $\chi_m -\chi_\perp$, where $\chi_m$ is the mean longitudinal in-plane susceptibility, shown in the range $T=150-330$ K. Solid lines are a linear fit. (d) Inverse of the temperature-dependent susceptibilities. Solid lines are a fit to the Curie-Weiss law above the structural transition $T_{S}$ (grey region). The extraction of the small transverse signal at high temperature leads to large systematic error $\gtrsim 250$ K (see text).}
\label{fig:CW_fit}
\end{figure}
%
The oscillation of the in-plane susceptibility is expected to vanish in the absence of bond anisotropies.
Furthermore, the difference $\chi_m-\chi_\perp$ (where $\chi_m=(\chi_+ + \chi_-)/2$ is the mean value of the in-plane oscillation) is proportional to the off-diagonal exchange couplings $\Gamma_1$ and $\Gamma_1'$,
\begin{equation}
\chi_\text{m}-\chi_\perp\sim T^{-2}(\Gamma_1'+2\Gamma_1)\;.
\end{equation}
That is, the observed anisotropy between in-plane and out-of-plane susceptibility originates from symmetric off-diagonal $\Gamma$ spin exchange.
As discussed below, a small $g$-factor anisotropy of the form  $g_x=g_y\neq g_z$ generates additional terms in these dependencies, however the overall trends remain unchanged.

As shown in Fig. 2, the experimentally observed oscillations of $\chi$ as a function of $\phi$, and the locations of their extrema, persist over large temperature ranges. Figure 3(a),(b) shows the temperature dependence of the longitudinal susceptibility measured perpendicular to the plane $\chi_\perp(T)$ and at the locations of the in-plane extrema $\chi_+(T)$ and $\chi_-(T)$, as well as the maximum transverse in-plane susceptibility $\chi_+^{\text{T}}(T)$. The temperature-dependent data were collected at fixed angle using standard, low-background sample holders to avoid diamagnetic contributions from the sample rotation stage. To confirm the validity of the high-temperature model, the differences $\chi_+-\chi_-$ and $\chi_\text{m}-\chi_\perp$ are shown in Fig. 3(c). The data plotted against $T^{-2}$ show reasonable correspondence with the linear behavior predicted by Eqn. (6) and (7).

\paragraph{Curie-Weiss analysis and model parameters.} The good agreement between the $\phi$-dependence of the experimentally measured susceptibility and the high-temperature expansion suggests a route to quantify the relationships between various model parameters. Due to symmetry considerations, the susceptibility tensor $\chi^{\mu\mu'}$ in Eq.~(\ref{chi_tensor}) has four independent components $\chi^{xx}$, $\chi^{zz}$, $\chi^{xy}$, $\chi^{xz}$, which allows the same number of exchange couplings to be determined. Since a bond-isotropic Kitaev model does not break the cubic symmetry of the interactions in spin space, it is generally impossible from susceptibility alone to distinguish between Heisenberg interactions $J_1$ and Kitaev exchange $K_1$ when fitting our experimental data to the high-temperature expansion. A possible set of linearly independent model parameters that can be determined in a fitting procedure is given by $\tilde{J}_1\equiv2J_1+J_1'+K_1$, $\Delta K_1\equiv K_1-K_1'$, $\Gamma_1$, $\Gamma_1'$.

Using the expansion in Eq.~(\ref{chi_tensor}), the inverse of the four susceptibility datasets shown in Fig. 3(a),(b) can be brought into the form $\chi^{-1}(T)\sim T-T_\text{CW}+\mathcal{O}(T^{-1})$ yielding four Curie-Weiss temperatures $T_{\text{CW}\perp}$, $T_{\text{CW}+}$, $T_{\text{CW}-}$, and $T_{\text{CW}+}^\text{T}$. These Curie-Weiss temperatures can be expressed as linear combinations of the model parameters. Defining the vectors $\mathcal{T}_\text{CW}=(T_{\text{CW}\perp},T_{\text{CW}+},T_{\text{CW}-},T_{\text{CW}+}^\text{T})$ and $\mathcal{J}=(\tilde{J}_1,\Delta K_1,\Gamma_1,\Gamma_1')$ one finds
$\mathcal{J}=k_\text{B}\mathcal{M}\mathcal{T}_\text{CW}$,
where $\mathcal{M}$ is a matrix which depends on the ratio $g_x/g_z$.

\begin{figure}[t]
\centering
\includegraphics[scale=0.5]{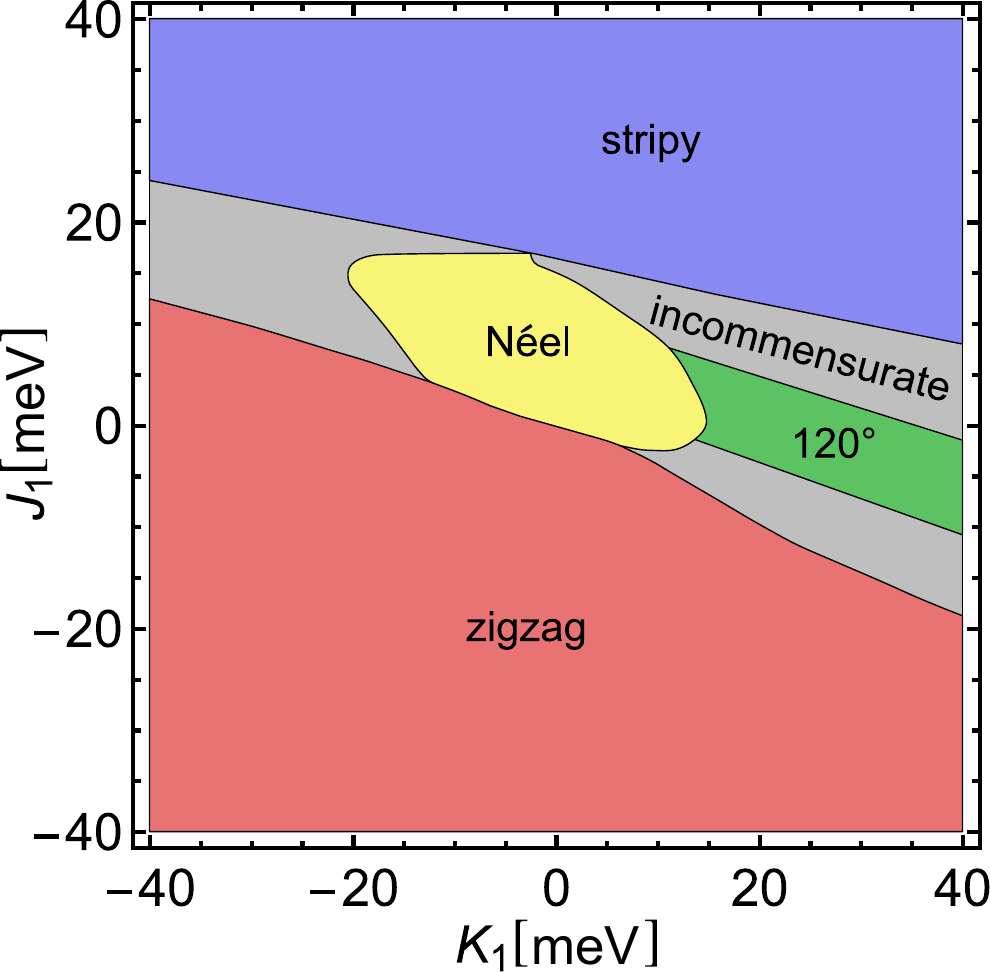}
\caption{Classical phase diagram of the anisotropic Hamiltonian in Eqs.~(\ref{ham1}) and (\ref{ham2}) as a function of $J_1$ and $K_1$ and fixed $2J_1+J_1'+K_1 = 14.3$\,meV, $K_1-K_1' = -7.7$\,meV, $\Gamma_1=29.8$\,meV, and $\Gamma_1'=27.9$\,meV.}
\label{phasediagram}
\end{figure}

The components of $\mathcal{T}_\text{CW}$ were determined by fitting a linear Curie-Weiss behavior to the high-temperature inverse susceptibilities $\chi_\perp^{-1}$, $\chi_+^{-1}$, $\chi_-^{-1}$, and $\chi_+^{\text{T}-1}$ (Fig.\, \ref{fig:CW_fit}d). The analysis is restricted to the high-temperature region 175 K $\leq T \leq$ 330 K away from the structural transition at $T_{S} \simeq $150 K \cite{park_emergence_2016} that produces kinks in the susceptibility curves. Fitting the longitudinal susceptibilities yields $T_{\text{CW}\perp}=-216.4(3)$ K, $T_{\text{CW}+}=39.6(2)$ K, and $T_{\text{CW}-}=32.6(3)$ K. At high temperatures, longitudinal contamination in the transverse SQUID coils is comparable to the intrinsic transverse signal, so that separating the two components introduces large errors (Fig. 3b,d). Therefore the Curie-Weiss fitting is performed over a narrower temperature range of 175 K $\leq T \leq$ 275 K to determine the intercept, $T_{\text{CW}+}^\text{T}=50(2)$ K. Based on these Curie-Weiss temperatures and the reported $g$-factor anisotropy of $g_x/g_z=1.1$ \cite{agrestini_electronically_2017} we obtain the model parameters $(\tilde{J}_1,\Delta K_1,\Gamma_1,\Gamma_1')=(14.3, -7.7, 29.8, 27.9)$ meV.

Inelastic neutron scattering \cite{banerjee2016,banerjee_neutron_2017,ran_spin-wave_2017,banerjee_excitations_2017,lampen-kelley_destabilization_2017} and most calculations \cite{janssen_magnetization_2017} place the magnetic exchange couplings for $\alpha$-RuCl$_{3}$ on the order of $\sim$5 - 10 meV, although $K_{1}$ as high as 16 meV \cite{do_majorana_2017} and recently 30 meV \cite{suzuki_strong_K} have also been proposed. The discrepancy in energy scale between lower estimates and the couplings of up to 30 meV in the model parameters determined above might be due to the
limited temperature ranges in which our Curie-Weiss fits are performed.
Despite the fact that our inverse susceptibility data are well described by a linear
behavior within our fitting range (see Fig.\,\ref{fig:CW_fit}), shifting the temperature
intervals upwards might still improve the results. Indeed, it has been argued for a Kitaev-Heisenberg model that depending on the precise
fitting range, experimentally determined Curie-Weiss temperatures need to be rescaled by factors of
2 or larger to obtain the true Curie-Weiss intercepts \cite{singh_high-temperature_2017}. We speculate that
such a rescaling (which in the simplest case would apply to all
interactions in the same way) would lead to exchange couplings with an overall size more consistent with other methods. Independent of such considerations, we conclude
that off-diagonal exchange $\Gamma$ and $\Gamma'$ plays a large role in the susceptibility of
$\alpha$-RuCl$_{3}$, in line with growing theoretical recognition of the importance of the $\Gamma$ term in the behavior of the system \cite{winter_challenges_2016, janssen_magnetization_2017}, including the recent prediction of a quantum spin liquid ground state in a Kitaev-$\Gamma$ model \cite{gohlke_signatures_2017}. Moreover, assuming the aforementioned model parameters and
mapping out the classical phase diagram within Luttinger-Tisza as a
function of the remaining free parameters $J_1$ and $K_1$, we indeed find the experimentally observed zigzag antiferromagnetic ground state in a large region of parameter space (see Fig.\,\ref{phasediagram}).

\paragraph{Conclusion.} 

The mapping out of the susceptibility tensor in single crystals of $\alpha$-RuCl$_3$ yields new insight into possibilities for the correct Hamiltonian describing the system. The phase shifts and periodicity of the observed in-plane oscillations can be understood within a bond-anisotropic spin Hamiltonian with substantial $\Gamma$ exchange. The agreement between the high-temperature expansion of the theoretical model and the measured oscillating susceptibilities $\chi_\parallel$ and $\chi^{\rm T}$ is remarkable, and indicates that the amplitude of the oscillations of susceptibility are proportional to the bond anisotropies in the Kitaev and $\Gamma$ terms. Our analysis further reveals that the marked easy-plane anisotropy in the system is a consequence of significant symmetric off-diagonal $\Gamma$ exchange.

\paragraph{Acknowledgements}
We acknowledge discussions with  J. van den Brink, B.Buechner, P.Gegenwart, L. Janssen, R.Thomale, M.Vojta, A.U.B. Wolter. P.L.K and D.M. were supported by the Gordon and Betty Moore Foundations EPiQS Initiative Grant GBMF4416. J.-Q.Y. and C.A.B. acknowledge support from the U.S. Department of Energy (US-DOE), Office of Science - Basic Energy Sciences (BES), Materials Sciences and Engineering Division. The work at the Oak Ridge National Laboratory High Flux Isotope Reactor was supported by US-DOE, Office of Science - BES, Scientific User Facilities Division. SR was supported by the DFG through SFB 1143. JR is supported by the Freie Universit\"at Berlin within the Excellence Initiative of the German
Research Foundation. 


\begin{thebibliography}{50}
\bibitem [{foo({\natexlab{a}})}]{footnote_2}%
  \BibitemOpen
\bibfield  {journal} {  }\href@noop {} {} ({\natexlab{a}}),\ \bibinfo {note}
  {it is worth emphasizing that the cartesian coordinates $x$, $y$, $z$
  appearing in the spin components of the Hamiltonian are different from the
  lattice directions ${\bf a}$, ${\bf b}$, ${\bf c}$ along which the
  susceptibility is measured. In particular, the direction perpendicular to the
  honeycomb plane is given by ${\bf c}=(\hat{x},\hat{y},\hat{z})/\sqrt{3}$ and
  the B1-bond is along the $(\hat{x},-\hat{y},0)/\sqrt{2}$ axis.}\BibitemShut
  {Stop}%
\end{thebibliography}

%

\end{document}